\documentclass[%
 reprint,
 amsmath,amssymb,
 aps,
prb,
]{revtex4-1}

\usepackage{graphicx}
\usepackage{dcolumn}
\usepackage{bm}

\allowdisplaybreaks[1]

\begin{document}

\preprint{APS/123-QED}

\title{Novel vortex structures in the three-dimensional superconductor under the helical magnetic field from the chiral helimagnet}

\author{Saoto Fukui$^1$}
\author{Masaru Kato$^1$}
\author{Yoshihiko Togawa$^2$}
\author{Osamu Sato$^3$}
\affiliation{$^1$Department of Mathematical Sciences, Osaka Prefecture University, 1-1, Gakuencho, Sakai, Osaka 599-8531, Japan}
\affiliation{$^2$Department of Physics and Electronics, Osaka Prefecture University, 1-1, Gakuencho, Sakai, Osaka 599-8531,Japan}
\affiliation{$^3$Osaka Prefecture University College of Technology, 26-12, Saiwaicho, Neyagawa, Osaka 572-8572, Japan}





\date{\today}

\begin{abstract}
We have investigated vortex structures in three-dimensional superconductors under a helical magnetic field from a chiral helimagnet numerically.
In order to obtain vortex structures, we solve three-dimensional Ginzburg-Landau equations with the finite element method.
The distribution of the helical magnetic field is assumed to be proportional to the distribution of the magnetic moments in the chiral helimagnet.
Then, the magnetic field is the same direction in the $yz$-plane and helical rotation along the helical axis.
Under this helical magnetic field, vortices appear to be perpendicular to the surface of the superconductor.
But we have found that there are tilted vortices toward the helical axis, although there is no component of the magnetic field along the helical axis.
This vortex structure depends on the chirality of the distribution of the helical magnetic field.
\end{abstract}


\maketitle


\section{Introduction}
\begin{figure}[t]
 \centering
 \includegraphics[scale=0.2]{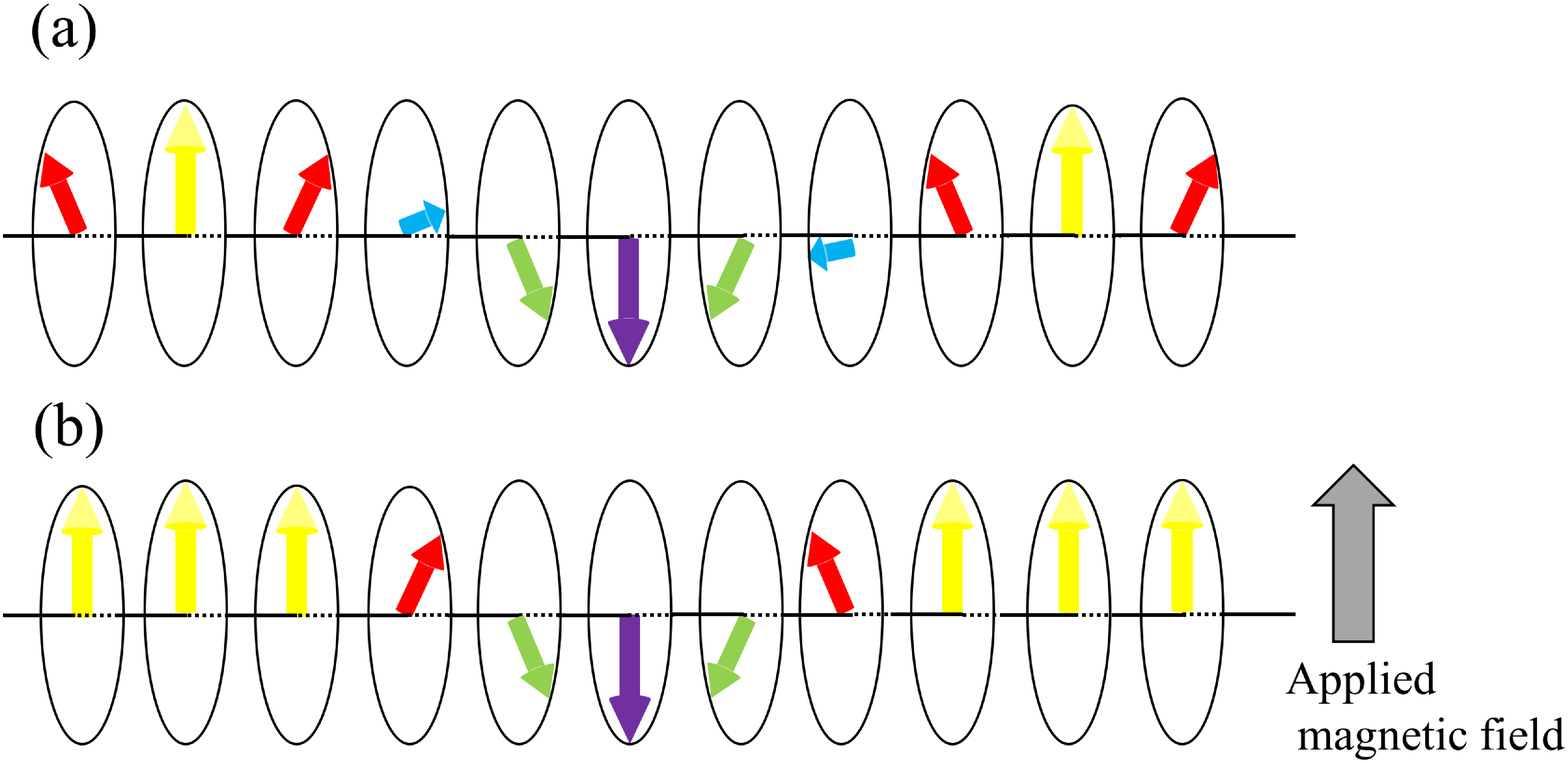}
 \caption{(a) Magnetic structure in the chiral helimagnet, and (b) the chiral soliton lattice under the applied magnetic field.}
 \label{chm_csl}
\end{figure}

A chiral helimagnet is one of the attractive magnetic materials.
In this material, magnetic moments form either left- or right-handed helical rotation.
The schematic magnetic structure of the chiral helimagnet is shown in Fig.~\ref{chm_csl}(a).
This helical configuration of magnetic moments comes from the competition between two interactions: the ferromagnetic exchange interaction between nearest neighbor magnetic moments and the Dzyaloshinsky-Moriya (DM) interaction.
The ferromagnetic exchange interaction causes nearest neighbor magnetic moments to be parallel, while the DM interaction causes nearest neighbor magnetic moments to be perpendicular to each other\cite{DM_1,DM_2}. 
The DM interaction determines the direction of the rotation, left-handed or right handed.
The magnetic structure in the chiral helimagnet has been observed in CrNb$_3$S$_6$\cite{CHM_Cr_1, CHM_Cr_2, CHM_Cr_3}, CsCuCl$_3$\cite{CHM_Cs_1, CHM_Cs_2}, and Yb(Ni$_{1-x}$Cu$_x$)$_3$Al$_9$\cite{CHM_Yb_1, CHM_Yb_2} experimentally.
Then, properties of chiral helimagnets have been investigated experimentally\cite{CHM_Cr_1, CHM_Cr_2, CHM_Cr_3,CHM_Cs_1, CHM_Cs_2,CHM_Yb_1,CHM_Yb_2} and theoretically\cite{CHM_theory_1, Kishine_theory}.

For example, under an external magnetic field perpendicular to the helical axis, the helical configuration of magnetic moments changes into an incommensurate magnetic structure\cite{Kishine_1}.
This magnetic structure is called a chiral soliton lattice, which is shown in Fig.~\ref{chm_csl}(b).
The chiral soliton lattice consists of ferromagnetic domains periodically partitioned by $360^\circ$ domain walls.
The chiral soliton lattice has been also observed experimentally with the Lorentz microscopy and the small-angle electron diffraction\cite{Togawa_PRL_1}.
Other interesting phenomena are the giant magnetoresistance\cite{Togawa_PRL_2}, the magneto-chiral dichroism\cite{MCh_Dh}, the creation of the spin current\cite{spin_current}, and the Berezinskii-Kosterlitz-Thouless (BKT) transition\cite{BKT}.

From other point of view, we may expect this peculiar magnetic structure affects other materials.
Therefore, in this paper, we focus on effects on superconductors, in particular, vortex structures in type-II superconductors.

These vortex structures in the type-II superconductor are important for a critical magnetic field and a critical current.
In general, under an uniform magnetic field, vortices appear and form a triangular lattice, which is called the Abrikosov lattice \cite{Abrikosov, Hess}.
When the external current flows in the superconductor and vortices move, the electric resistance occurs, which leads to break superconductivity.
So, controlling vortex states is a key factor for applications of superconductors.

One of plausible vortex controlling method is using a ferromagnet.
In a ferromagnet / superconductor bilayer system, vortices appear spontaneously \cite{FM_SC_Hybrid, FM_SC_FSB_ex, FM_SC_FSB_theory}.
This phenomenon comes from interaction between magnetic domains in the ferromagnet and magnetic fluxes of vortices.
This spontaneous vortex state enhances superconductivity, in particular, its critical current by a pinning of vortices in the superconductor due to magnetic domains in the ferromagnet. 
From this interference between the ferromagnet and the superconductor, we expect novel effects of peculiar magnetic materials; the chiral helimagnet.
Therefore, we investigate effects of the chiral helimagnet on the superconductor theoretically.

In our previous study, we have investigated vortex structures in two-dimensional superconductors under the magnetic field from the chiral helimagnet\cite{Fukui_SUST, Fukui_JPSJ}.
Although, magnetic field is created by the helically structured magnetic moments in the chiral helimagnet only the perpendicular component of the magnetic field to the superconducting surface is effective, and other components are neglected in the two-dimensional superconductor, .
In this paper, we consider three-dimensional superconductors, and taking all components of the magnetic field, investigate vortex structures under the helical magnetic field from the chiral helimagnet completely.
In section II, we introduce numerical methods in order to obtain vortex structures in three-dimensional superconductors.
In section III, we show vortex structures under the helical magnetic field from the chiral helimagnet, and discuss origin of these vortex structures.
Finally, in section IV, we summarize our results.

\section{Method}
We consider a three-dimensional superconductor under the helical magnetic field.
The distribution of the helical magnetic field is assumed to be proportional to the distribution of the magnetic moments in the chiral helimagnet.
We obtain distributions of the order parameter in superconductors by solving the Ginzburg-Landau equations.
We start from the Ginzburg-Landau free energy,
\begin{eqnarray}
 & &\mathcal{F}(\psi,\mbox{\boldmath $A$}) = \int_V \left( f_n + \alpha(T) |\psi|^2 + \frac{\beta}{2} |\psi|^4 \right) dV \nonumber \\
                                                                & &  + \int_V \left\{ \frac{1}{2m_s} \left| \left( -i\hbar \nabla - \frac{e_s \mbox{\boldmath $A$}}{c} \right) \psi \right|^2 
                                                                 + \frac{|\mbox{\boldmath $h$}|^2}{8\pi} - \frac{\mbox{\boldmath $h$} \cdot \mbox{\boldmath $H$}_{\rm ext}}{4\pi} \right\} dV, \nonumber \\ \label{gl_free_3d}
\end{eqnarray}
where $\psi$ is a superconducting order parameter and $f_n$ is a free energy density of the normal state.
$\alpha(T)$ is a coefficient which depends on the temperature $T$, $\alpha(T) = \alpha'(T-T_c)$.
$\alpha'$ and $\beta$ is a positive constant and $T_c$ is the critical temperature of the superconductor.
$m_s$ is the effective mass of the superconductor and $e_s$ is an effective charge of electrons in the superconductor.
$\mbox{\boldmath $h$} = \nabla \times \mbox{\boldmath $A$}$ is a local magnetic field and $\mbox{\boldmath $A$}$ is a magnetic vector potential.
$\mbox{\boldmath $H$}_{\rm ext}$ is an external magnetic field, which is included the magnetic field from the chiral helimagnet.
The order parameter and the vector potential is normalized as,
\begin{equation}
 \tilde{\psi} = \frac{\psi}{\sqrt{\alpha}/\beta},~~~\tilde{\mbox{\boldmath $A$}} = \frac{2\pi}{\Phi_0} \mbox{\boldmath $A$}. \label{normalize}
\end{equation}
$\Phi_0 = ch/2e$ is the quantum flux and $e$ is an elementary charge.
Using the normalized order parameter and the vector potential, we obtain following equations from Eq.~(\ref{gl_free_3d}),
\begin{eqnarray}
& & \int_{V} \left[ \left( i\nabla \tilde{\psi} - \tilde{\mbox{\boldmath $A$}}\tilde{\psi} \right) \left( -i \nabla (\delta \tilde{\psi}) - \tilde{\mbox{\boldmath $A$}} (\delta \tilde{\psi}) \right) \right.  \nonumber \\ 
& & \left. + \left( i\nabla (\delta\tilde{\psi}) - \tilde{\mbox{\boldmath $A$}} (\delta \tilde{\psi}) \right) \left( -i\nabla \tilde{\psi}^\ast - \tilde{\mbox{\boldmath $A$}} \tilde{\psi}^\ast \right) \right. \nonumber \\
& & \left. + \frac{1}{\xi^2} \left( |\tilde{\psi}|^2 - 1 \right) \left( \tilde{\psi} (\delta \tilde{\psi}^\ast) + \tilde{\psi}^\ast (\delta \tilde{\psi}) \right) \right] d\Omega = 0, \label{gl_3d_1} \\ \nonumber \\
& & \int_{V} \left[ \kappa^2 \xi^2 \left( {\rm div}~\tilde{\mbox{\boldmath $A$}} \cdot {\rm div}~(\delta \tilde{\mbox{\boldmath $A$}}) + {\rm rot}~\tilde{\mbox{\boldmath $A$}} \cdot {\rm rot}~(\delta \tilde{\mbox{\boldmath $A$}}) \right) \right. \nonumber \\
& & \left. |\psi|^2 \tilde{\mbox{\boldmath $A$}} \cdot (\delta \tilde{\mbox{\boldmath $A$}}) - \frac{i}{2} \left\{ \tilde{\psi}^\ast (\nabla \tilde{\psi}) - \tilde{\psi} (\nabla \tilde{\psi}^\ast) \right\} \tilde{\mbox{\boldmath $A$}} \right] d\Omega \nonumber \\
& & = \kappa^2 \xi^2 \int_V \frac{2\pi}{\Phi_0} \mbox{\boldmath $H$}_{\rm ext} \cdot {\rm rot}~(\delta \tilde{\mbox{\boldmath $A$}}) d\Omega, \label{gl_3d_2}
\end{eqnarray}
where $\delta \tilde{\psi}$ and $\delta\tilde{\mbox{\boldmath $A$}}$ are variations, or test functions of the order parameter and the vector potential, respectively.
$\kappa = \lambda/\xi$ is the Ginzburg-Landau parameter, and $\lambda$ and $\xi$ are the penetration length and the coherence length, respectively.

\begin{figure}[htbp]
 \centering
 \includegraphics[scale=0.3]{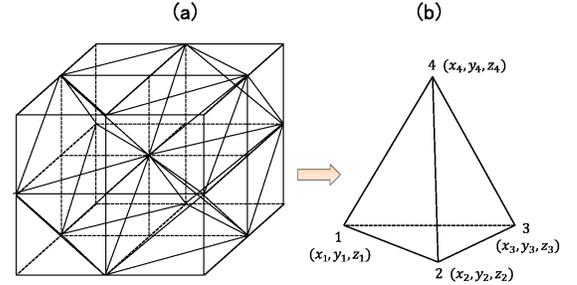}
 \caption{(a) Schematic three-dimensional system, (b) a tetrahedron finite element. Coordinates of four nodes $i=1,~2,~3,~4$ of the $e$-th tetrahedron denote $1(x_1,~y_1,~z_1)$, $2(x_2,~y_2,~z_2)$, $3(x_3,~y_3,~z_3)$, and $4(x_4,~y_4,~z_4)$. }
 \label{system_finite_element}
\end{figure}

In order to solve Eqs.~(\ref{gl_3d_1}) and (\ref{gl_3d_2}), we use the three-dimensional finite element method (FEM).
In the three-dimensional FEM, we divide the system into tetrahedron elements (see Fig.~\ref{system_finite_element}).
In a tetrahedron, there are four volume coordinates, which is given as,
\begin{equation}
 N_i^e = \left( a_i + b_ix + c_i + d_iz \right)/6V~~~~(i=1,~2,~3,~4), \label{finite_element_3d}
\end{equation}
where $V$ is a volume of the tetrahedron.
$a_i,~b_i,~c_i,$ and $d_i$ are given in the Appendix.
The order parameter $\tilde{\psi}$ and the vector potential $\tilde{\mbox{\boldmath $A$}}$ are expanded using these volume coordinates,
\begin{eqnarray}
 \tilde{\psi}(\mbox{\boldmath $x$}) &=& \sum_{i,e}^{N_e} \tilde{\psi}_i^e N_i^e, \label{psi_expand} \\
 \tilde{\mbox{\boldmath $A$}} (\mbox{\boldmath $x$}) &=& \sum_{i,e}^{N_e} \tilde{\mbox{\boldmath $A$}}_i^e N_i^e, \label{a_expand}
\end{eqnarray} 
where $\tilde{\psi}_i^e$ and $\tilde{\mbox{\boldmath $A$}}_i^e$ are the order parameter and the vector potential at $i-$th node in the $e-$th element, respectively.
$N_e$ is the number of finite elements.
We consider following test functions $\delta \tilde{\psi}$ and $\delta \tilde{\mbox{\boldmath $A$}}$,
\begin{eqnarray}
 \delta \tilde{\psi}_{8(e-1) + 2j-1} &=& \begin{cases} N_j^e~~~~(x \in e\text{-th~element})  \\ 0 ~~~~~~~(\text{otherwise}), \end{cases} \label{test_psi_1} \\
 \delta \tilde{\psi}_{8(e-1) + 2j}    &=& \begin{cases} iN_j^e~~~(x \in e\text{-th~element}) \\ 0 ~~~~~~~(\text{otherwise}), \end{cases} \label{test_psi_2} \\ 
 \delta \tilde{\mbox{\boldmath $A$}}_{12(e-1) + 3(j-1)+i} &=& \begin{cases} \mbox{\boldmath $e$}_i N_j^e~(x \in e\text{-th~element})  \\ 0 ~~~~~~~(\text{otherwise}), \end{cases} \label{test_a} 
\end{eqnarray}
where $i=1,2,3$ and $j=1,2,3,4$.
$\mbox{\boldmath $e$}_1,~\mbox{\boldmath $e$}_2, $ and $\mbox{\boldmath $e$}_3$ are basis vectors in the three-dimensional space.
We substitute Eqs. (\ref{psi_expand})-(\ref{test_a}) into Eqs. (\ref{gl_3d_1}) and (\ref{gl_3d_2}), we obtain these equations,
\begin{eqnarray}
 & & \sum_j \left[ P_{ij}(\{\tilde{A} \}) + P_{ij}^{2R}(\{ \tilde{\psi} \}) \right] {\rm Re} \tilde{\psi}_j^e \nonumber \\
 & & ~ + \sum_j \left[ Q_{ij} (\{ \tilde{\mbox{\boldmath $A$}} \}) + Q_{ij}^{2}(\{ \tilde{\psi} \}) \right] {\rm Im} \tilde{\psi}_j^e = V_i^{R}(\{ \tilde{\psi} \}),    \label{e1} \\
 & & \sum_j \left[ -Q_{ij}(\{\tilde{\mbox{\boldmath $A$}}\}) + Q_{ij}^{2}(\{\tilde{\psi}\}) \right] {\rm Re} \tilde{\psi}_j^e \nonumber \\
 & & ~ + \sum_j \left[ P_{ij}(\{\tilde{\mbox{\boldmath $A$}}\}) + P_{ij}^{2I}(\{\tilde{\psi}\}) \right] {\rm Im} \tilde{\psi}_j^e = V_{i}^{I}(\{\tilde{\psi}\}), \label{e2} \\
 & & \sum_j R_{ij} (\{\tilde{\psi}\}) \tilde{A}_{jx} + \sum_j S_{ij}^{xy} \tilde{A}_{jy} + \sum_j S_{ij}^{xz} \tilde{A}_{jz} \nonumber \\
 & & ~ = T_i^{x} - U_i^{x}, \label{e3} \\
 & & \sum_j R_{ij} (\{\tilde{\psi}\}) \tilde{A}_{jy} + \sum_j S_{ij}^{yx} \tilde{A}_{jx} + \sum_j S_{ij}^{yz} \tilde{A}_{jz} \nonumber \\
 & & ~ = T_i^{y} - U_i^{y}, \label{e4} \\
 & & \sum_j R_{ij} (\{\tilde{\psi}\}) \tilde{A}_{jz} + \sum_j S_{ij}^{zx} \tilde{A}_{jx} + \sum_j S_{ij}^{zy} \tilde{A}_{jy} \nonumber \\
 & & ~ = T_j^{z} - U_i^{z}. \label{e5}
\end{eqnarray}
Coefficients are given in Appendix and the reference \cite{Fukui_3d_proc}.
Solving Eqs.~(\ref{e1})-(\ref{e5}) self consistently, we obtain real and imaginary parts of the order parameter ${\rm Re}~\tilde{\psi}$, ${\rm Im}~\tilde{\psi}$ and three-components of the vector potential $A_x,~A_y,$ and $A_z$.

The magnetic field $\mbox{\boldmath $H$}_{\rm ext}$ includes the helical magnetic field from the chiral helimagnet, 
\begin{equation}
 \mbox{\boldmath $H$}_{\rm ext} = \mbox{\boldmath $H$}_{\rm CHM} + \mbox{\boldmath $H$}_{\rm appl}, \label{external_h}
\end{equation}
where $\mbox{\boldmath $H$}_{\rm CHM}$ is the magnetic field from the chiral helimagnet and $\mbox{\boldmath $H$}_{\rm appl}$ is the homogeneous applied magnetic field.
We consider that a distribution of $\mbox{\boldmath $H$}_{\rm CHM}$ is proportional to the configuration of magnetic moments in the chiral helimagnet.
The configuration of magnetic moments is obtained by the Hamiltonian of the chiral helimagnet \cite{Kishine_1, Fukui_SUST, Fukui_JPSJ, Fukui_3d_proc},
\begin{eqnarray}
 \mathcal{H} &=& -2J \sum_{n} \mbox{\boldmath $S$}_n \cdot \mbox{\boldmath $S$}_{n+1} + \mbox{\boldmath $D$} \cdot \sum_n \mbox{\boldmath $S$}_n \times \mbox{\boldmath $S$}_{n+1} \nonumber \\
                     & & - 2\mu_B \mbox{\boldmath $H$}_{\rm appl} \cdot \sum_n \mbox{\boldmath $S$}_n, \label{hamiltonian_chm}
\end{eqnarray}
where $\mbox{\boldmath $S$}_n$ is the spin at the $n$-th site, $\mu_B$ is the Bohr magneton.
This Hamiltonian consists of three terms.
The first term is the ferromagnetic exchange interaction with magnitude $J~(>0)$.
The second term is the DM interaction with the DM vector $\mbox{\boldmath $D$}$.
The last term is the Zeeman energy.
We assume that the helical axis is the $x$-axis.
In the monoaxial chiral helimagnet, the DM vector is parallel to the helical axis, $\mbox{\boldmath $D$} = (D,0,0)$, and we assume that the direction of spin is perpendicular to the helical axis, so $\varphi = \pi/2$.
We express $n$-th spin as $\mbox{\boldmath $S$}_n = S(\sin{\theta_n} \cos{\varphi},~\sin{\theta_n} \sin{\varphi},~\cos{\theta_n})$.
We set $\mbox{\boldmath $H$}_{\rm appl} = (0,0,H_{\rm appl})$.
In the typical chiral helimagnet CrNb$_3$S$_6$, the helical period $L=48$nm is much longer than the lattice constant.
So, we consider the continuum limit.
We minimize Eq. (\ref{hamiltonian_chm}) in the continuum limit with respect to $\theta(x)$.
We obtain the Sine-Gordon equation,
\begin{equation}
 \frac{d^2\theta(x)}{dx^2} - H^\ast \sin{\theta(x)} = 0, \label{sine_gordon}
\end{equation}
where $H^\ast = 2\mu_B H_{\rm appl}/(a^2S^2\sqrt{J^2 + |\mbox{\boldmath $D$}|^2})$ is a normalized applied magnetic field and $a$ is the lattice constant.
The solution of Eq.(\ref{sine_gordon}) is,
\begin{equation}
 \sin{\left( \frac{\theta - \phi}{2} \right)} = {\rm sn} \left( \frac{\sqrt{H^\ast}}{k}x~|~k \right), \label{theta_1}
\end{equation}
or,
\begin{equation}
 \theta(x) = 2\sin^{-1} \left[ {\rm sn} \left( \frac{\sqrt{H^\ast}}{k}x~|~k \right) \right] + \phi. \label{theta_2}
\end{equation}
${\rm sn}(x|k)$ is the Jacobi's elliptic function, $k$ is the modulus and $\phi$ is an initial angle at $x=0$.
$k$ is determined by the relation,
\begin{equation}
 \frac{\pi \tan^{-1}(|\mbox{\boldmath $D$}|/J)}{4\sqrt{H^\ast}} = \frac{E(k)}{k}, \label{k_det}
\end{equation}
where $E(k)$ is the complete elliptic integral of the second kind.
Using Eq. (\ref{theta_2}), the external magnetic field in Eq. (\ref{external_h}) is,
\begin{eqnarray}
 (\mbox{\boldmath $H$}_{\rm ext})_x(x) &=& 0, \label{h_x} \\
 (\mbox{\boldmath $H$}_{\rm ext})_y(x) &=& H_0 \sin{\theta(x)}, \label{h_y} \\
 (\mbox{\boldmath $H$}_{\rm ext})_z(x) &=& H_0 \cos{\theta(x)} + H_{\rm appl}. \label{h_z}
\end{eqnarray}
$H_0$ is a magnitude of the magnetic field from the chiral helimagnet.
We solve the Ginzburg-Landau equations (\ref{e1})-(\ref{e5}) using the magnetic field in Eqs. (\ref{h_x})-(\ref{h_z}) numerically.
\section{Result}
\begin{figure}[t]
 \centering
 \includegraphics[scale=0.3]{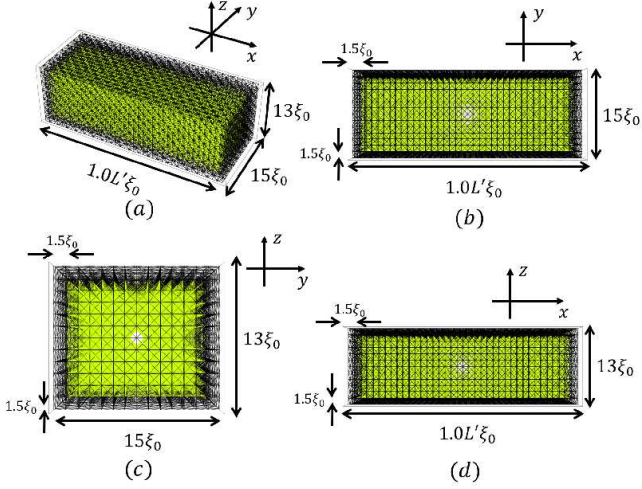}
 \caption{(a) Three-dimensional superconductor with finite tetrahedron elements, (b) $xy$-plane of the system, (c) $yz$-plane of the system, (d) $zx$-plane of the system. The system size in these figures is $1.0L'\xi_0 \times 15\xi_0 \times 13\xi_0$. The superconducting region (Green) is surrounded by the vacuum region (Black). The distance between the superconducting region is the vacuum region is $1.5\xi_0$.}
 \label{Fig_system}
\end{figure}

Solving the Ginzburg-Landau equations (\ref{e1})-(\ref{e5}) self-consistently, we obtain distributions of the order parameter in the superconductor under the chiral helimagnet.
We set the Ginzburg-Landau parameter $\kappa = \lambda/\xi = 5$ and the temperature $T=0.3T_c$, where $T_c$ is the critical temperature of the superconductor.
The ratio between the ferromagnetic exchange interaction and the Dzyaloshinsky-Moriya interaction is taken from the experimental data\cite{DM_Cr}, $|\mbox{\boldmath $D$}|/J = 0.16$.
We consider a parallelepiped as our model shown in Fig.~\ref{Fig_system}.
The system size is $1.0L'\xi_0 \times 15\xi_0 \times 13\xi_0$.
$\xi_0$ is a coherence length at $T=0$ and $L'\xi_0$ is a helical period of the chiral helimagnet, which is given as,
\begin{equation}
 L' = \frac{2\pi}{\tan^{-1}(D/J)}. \label{period}
\end{equation}
Here, the uniform applied magnetic field, $H_{\rm appl}/(\Phi_0/\xi_0^2) = 0.00$.
For $|\mbox{\boldmath $D$}|/J=0.16$, $L'$ is approximately $39.2699$.
We assume the superconducting region is surrounded by the vacuum region.
The distance between the superconducting region and the vacuum region is $1.5\xi_0$, and the size of the superconducting region is $(1.0L' - 3.0)\xi_0 \times 12\xi_0 \times 10\xi_0$.
When we calculate the Ginzburg-Landau equations, we set following boundary conditions,
\begin{equation} 
 \mbox{\boldmath $A$} \cdot \mbox{\boldmath $n$} = 0,~~\left| \left( -i\hbar \nabla + \frac{e\mbox{\boldmath $A$}}{c} \right) \psi \right| \cdot \mbox{\boldmath $n$} = 0. \label{boundary}
\end{equation}

\begin{figure}[t]
 \centering
 \includegraphics[scale=0.27]{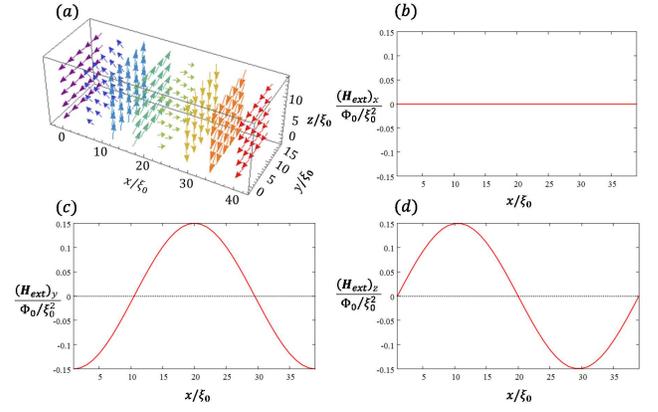}
 \caption{(a) Distributions of the magnetic field from the chiral helimagnet, (b) $x$-component of the magnetic field, (c) $y$-component of the magnetic field, (d) $z$-component of the magnetic field. The amplitude of the helical magnetic field is $H_0/(\Phi_0/\xi_0^2) = 0.15$ and the applied magnetic field is $H_{\rm appl}/(\Phi_0/\xi_0^2) = 0.00$. }
 \label{Fig_field_1}
\end{figure}

\begin{figure}[htbp]
 \centering
 \includegraphics[scale=0.3]{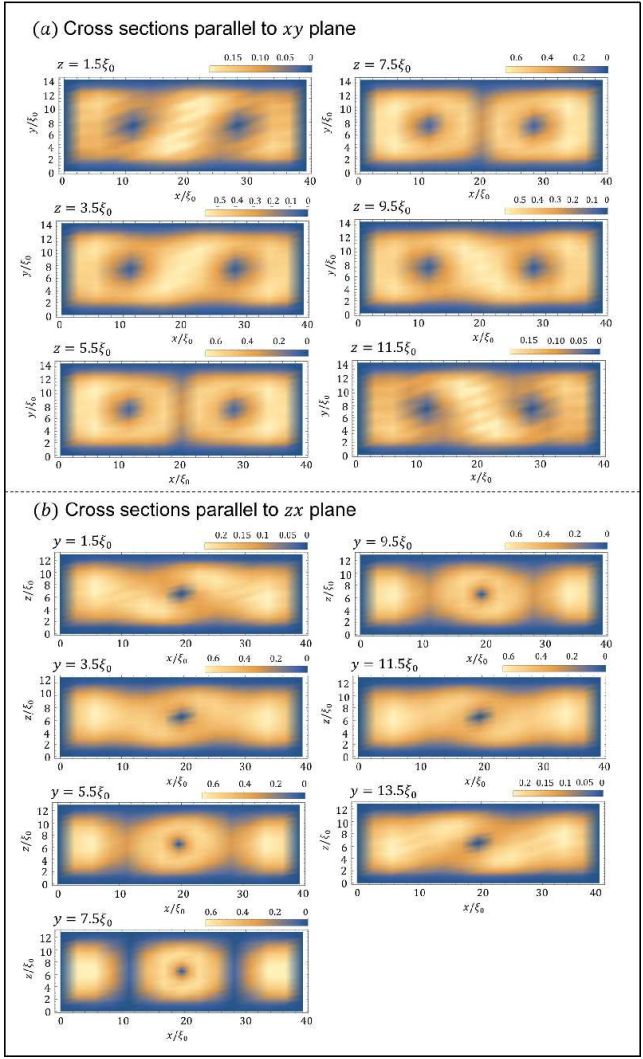}
 \caption{Distributions of the order parameter in cross sections parallel to (a) $xy$-plane and (b) $zx$-plane. The amplitude of the helical magnetic field $H_0/(\Phi_0/\xi_0^2) = 0.15$ and the applied magnetic field $H_{\rm appl}/(\Phi_0/\xi_0^2) = 0.00$. }
 \label{op_1_2}
\end{figure}

The external magnetic field $\mbox{\boldmath $H$}_{\rm ext}$ is given in Eqs.~(\ref{h_x})-(\ref{h_z}).
We set magnitudes of the helical magnetic field and the applied magnetic field in Eqs.~(\ref{h_x})-(\ref{h_z}) as $H_0/(\Phi_0/\xi_0^2) = 0.15$ and $H_{\rm appl}/(\Phi_0/\xi_0^2) = 0.00$, respectively.
First, we take the angle at $x=0$ in Eqs. (\ref{theta_1}) or (\ref{theta_2}) as $\phi = -\pi/2$.
Then, the distribution of the helical magnetic field is shown in Fig.~\ref{Fig_field_1}.
In Fig.~\ref{Fig_field_1}, we show distribution of the helical magnetic field (Fig.~\ref{Fig_field_1}(a)), each components of the magnetic field $(\mbox{\boldmath $H$}_{\rm ext})_x$, $(\mbox{\boldmath $H$}_{\rm ext})_y$, and $(\mbox{\boldmath $H$}_{\rm ext})_z$ (Figs.~\ref{Fig_field_1}(b)-(d)).
Under the magnetic field in Fig.~\ref{Fig_field_1}, we obtain the distribution of the order parameter shown in Fig.~\ref{op_1_2}.
In Fig.~\ref{op_1_2}, we show distributions of the order parameter in the $xy$-plane (Fig.~\ref{op_1_2}(a)) and the $zx$-plane (Fig.~\ref{op_1_2}(b)).
The cross sections parallel to the $xy$-plane at $z=1.5\xi_0,~11.5\xi_0$ and the $zx$-plane at $y=1.5\xi_0,~13.5\xi_0$ are interfaces between the superconducting region and the vacuum region.
In Fig.~\ref{op_1_2}(a), two vortices appear in the regions around $(x/\xi_0,~y/\xi_0) \sim (10,~7.5)$ and $(30,~7.5)$, where $(\mbox{\boldmath $H$}_{\rm ext})_y/(\Phi_0/\xi_0^2) \sim 0.00$ and $(\mbox{\boldmath $H$}_{\rm ext})_z/(\Phi_0/\xi_0^2) \sim \pm 0.15$ in Fig.~\ref{Fig_field_1}.
In this regions, the magnetic field is parallel or anti-parallel to $z$-axis, so these two vortices have quantum fluxes antiparallel to each other.
On the other hand, in Fig.~\ref{op_1_2}(b), one vortex appears in the region around $(x/\xi_0,~y/\xi_0) \sim (20,~7.5)$, where $(\mbox{\boldmath $H$}_{\rm ext})_y/(\Phi_0/\xi_0^2) \sim 0.15$ and $(\mbox{\boldmath $H$}_{\rm ext})_z/(\Phi_0/\xi_0^2) \sim 0.00$.
In this region, the magnetic field is parallel to the direction of $y-$axis, so the vortex has a quantum flux parallel to the $y-$axis.
In total, three vortices appear.
They are separated by $0.25\xi_0$.
And the angle between nearest neighbor vortices is $\pi/2$. 

The vortex structure is different from our previous work\cite{Fukui_3d_proc}, although the model and numerical parameters are the same.
Only difference is the initial random state of our iteration method for solving Eqs.~(\ref{e1})-(\ref{e5}).
To determine more stable state, we should calculate the free energies for both states.
This is the future problem.

\begin{figure}[t]
 \centering
 \includegraphics[scale=0.27]{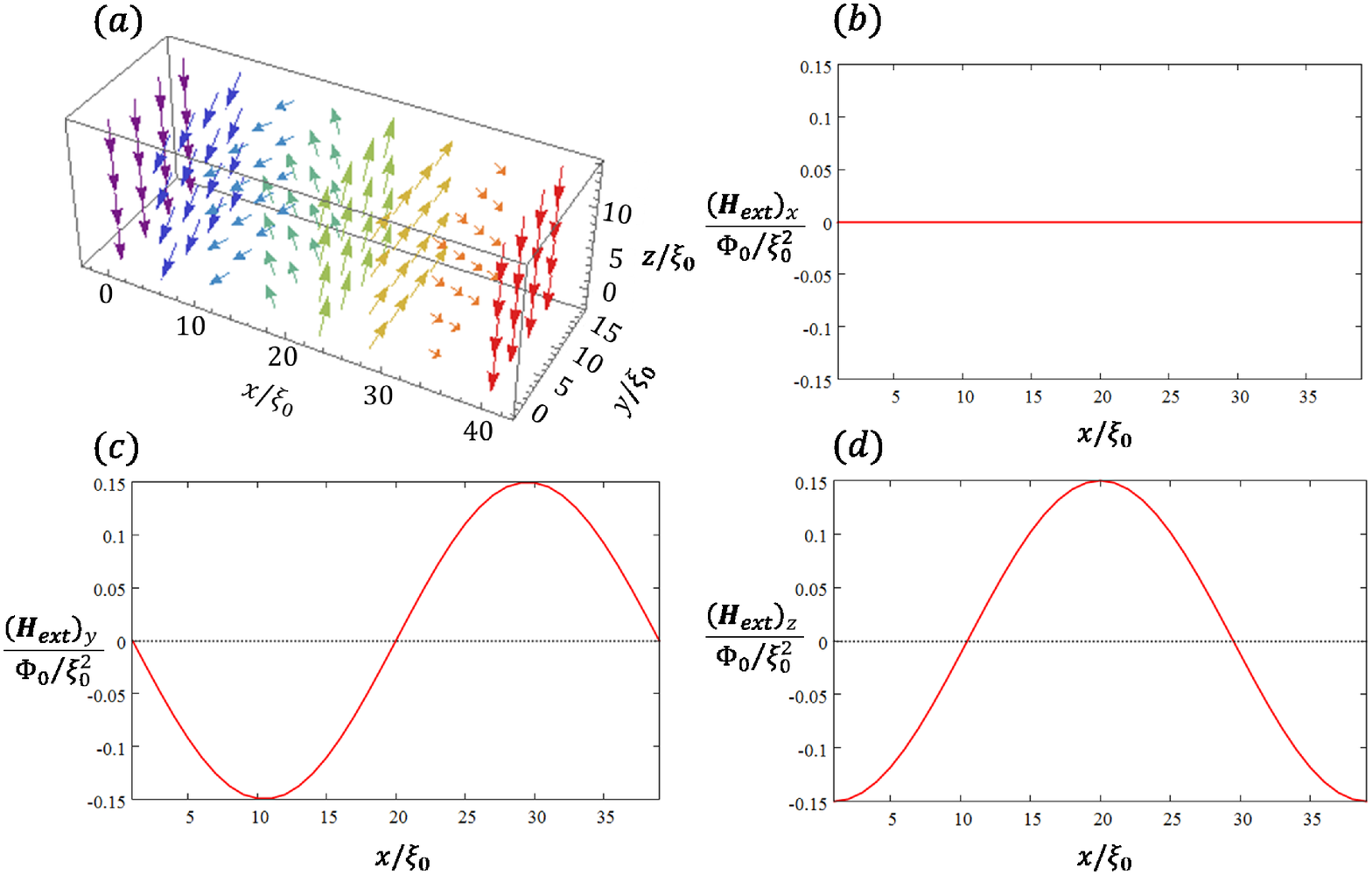}
 \caption{(a) Distributions of the magnetic field from the chiral helimagnet, (b) $x$-component of the magnetic field, (c) $y$-component of the magnetic field, (d) $z$-component of the magnetic field. The amplitude of the helical magnetic field is $H_0/(\Phi_0/\xi_0^2) = 0.15$ and the applied magnetic field is $H_{\rm appl}/(\Phi_0/\xi_0^2) = 0.00$ for $\phi = \pi$ in Eq.~(\ref{theta_2}). }
 \label{Fig_field_2}
\end{figure}

\begin{figure*}[htbp]
 \centering
 \includegraphics[scale=0.32]{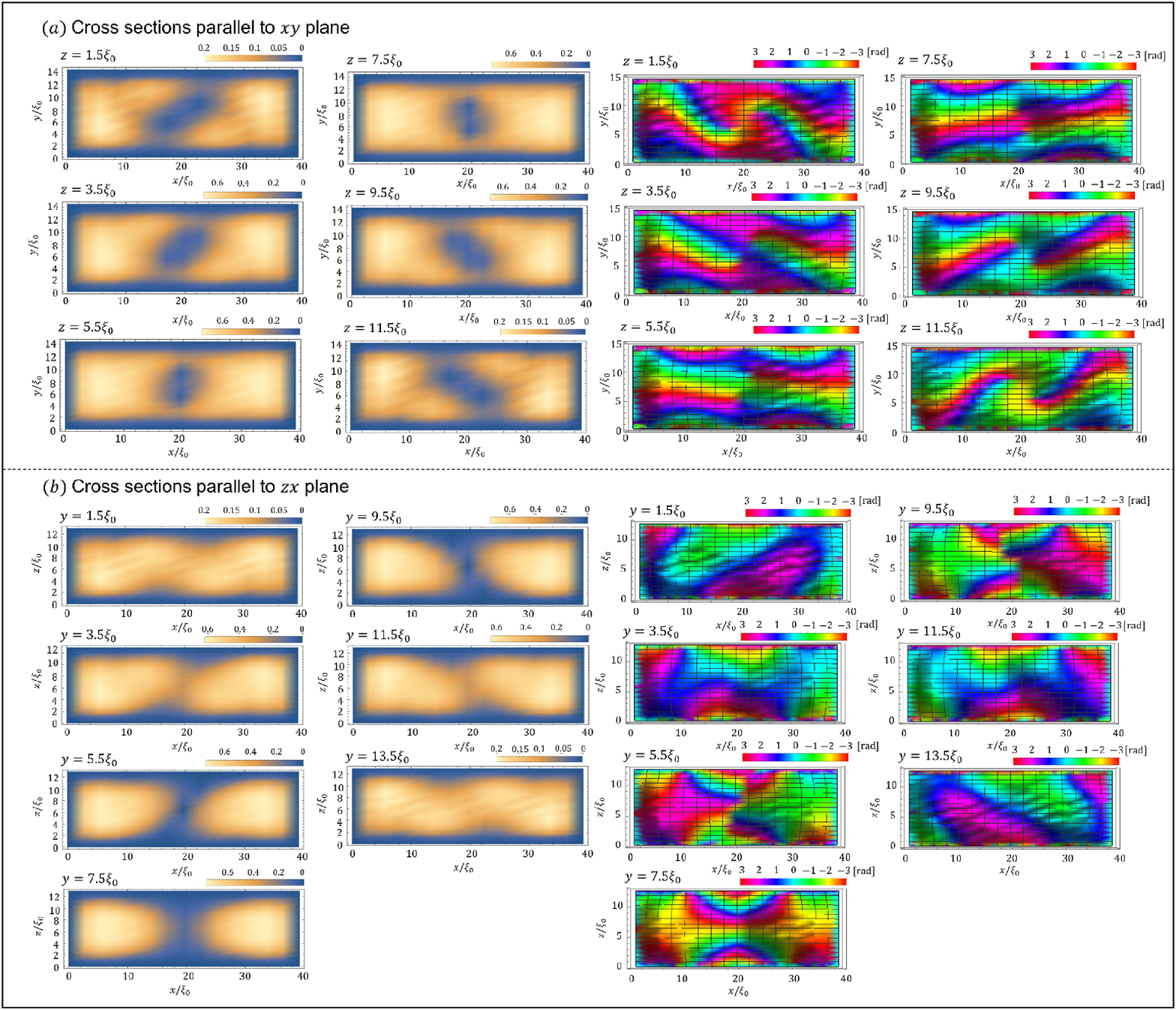}
 \caption{Distributions of the order parameter and phases of the order parameter in the cross sections parallel to (a) $xy$-planes and (b) $zx$-planes. The amplitude of the helical magnetic field is $H_0/(\Phi_0/\xi_0^2) = 0.15$ and the applied magnetic field is $H_{\rm appl}/(\Phi_0/\xi_0^2) = 0.00$. }
 \label{op_2_2}
\end{figure*}

\begin{figure*}[htbp]
 \centering
 \includegraphics[scale=0.3]{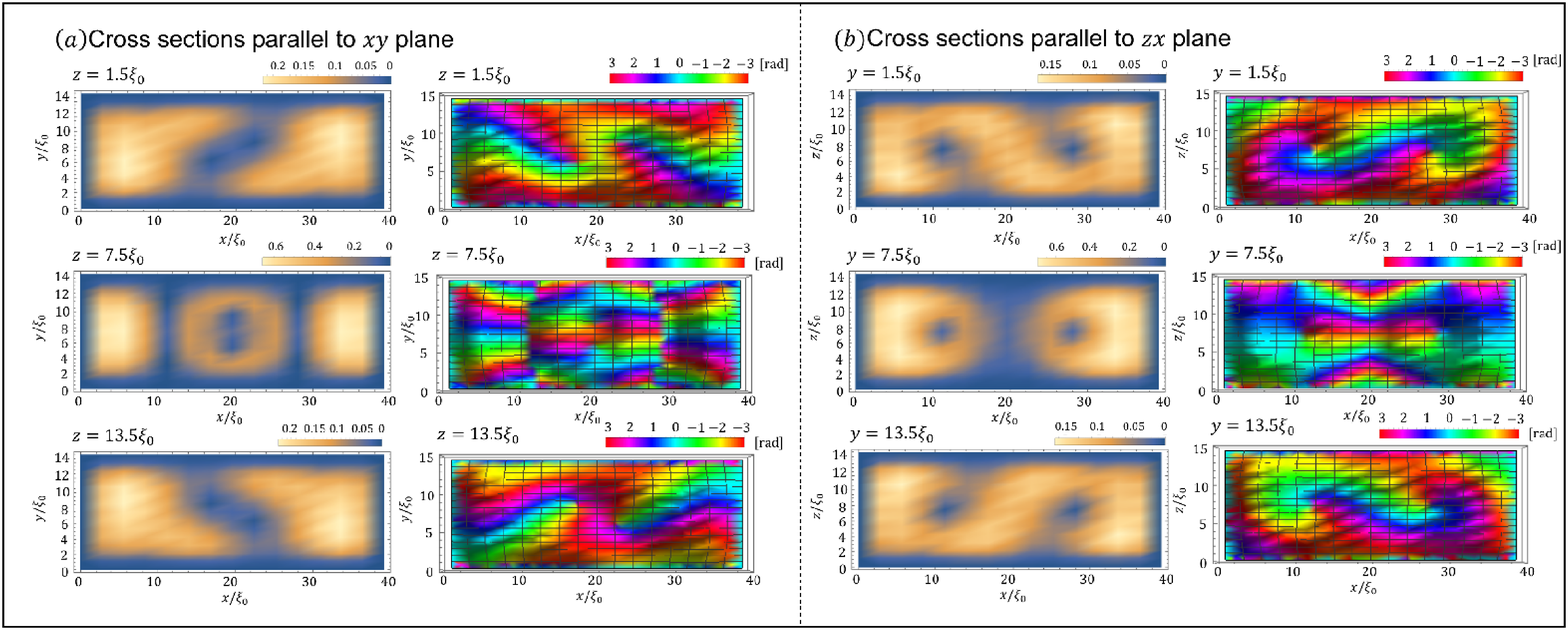}
 \caption{Distributions of the order parameter and phases of the order parameter in the cross sections parallel to (a) $xy$-planes and (b) $zx$-planes. The amplitude of the helical magnetic field $H_0/(\Phi_0/\xi_0^2) = 0.15$ and the applied magnetic field $H_{\rm appl}/(\Phi_0/\xi_0^2) = 0.00$. }
 \label{op_3_2}
\end{figure*}

Next, we investigate vortex structures with the other distribution of the helical magnetic field with $\phi = \pi$ in Eq.~(\ref{theta_1}), which is shown in Fig,~\ref{Fig_field_2}.
Under this magnetic field, we obtain distributions of the order parameter shown in Fig.~\ref{op_2_2}.
In Fig.~\ref{op_2_2}, we show the distributions of the order parameter and the phases of the order parameter.
In Fig. \ref{op_2_2}(a), two vortices appear in the region of the magnetic field $(\mbox{\boldmath $H$}_{\rm ext})_z/(\Phi_0/\xi_0) > 0$.
Positions of these vortices in these cross sections from $z=1.5\xi_0$ to $11.5\xi_0$ change along the $x$-axis.
So, two vortices tilt toward the $x$-axis, but the $x$-component magnetic field is zero, $(\mbox{\boldmath $H$}_{\rm ext})_x/(\Phi_0/\xi_0^2) = 0$.
Vortices parallel to the $y$-axis does not appear in Fig.~\ref{op_2_2}(b). 
This result comes from the screening current of demagnetization factor of the superconductor.
Next, in order to avoid difference between shielding fields parallel to $y$- and $z$-axis, we investigate vortex structure in the system with the larger size.
The system size is $1.0L'\xi_0 \times 15\xi_0 \times 15\xi_0$.
In this system, the cross section parallel to the $yz$-plane is square.
We take the same numerical parameters $\kappa = 5,~T=0.3T_c,$ and $|\mbox{\boldmath $D$}|/\xi_0 = 0.16$, and the same distribution of the helical magnetic field, $H_0/(\Phi_0/\xi_0^2) = 0.15,~H_{\rm appl}/(\Phi_0/\xi_0^2) = 0.0$, and $\phi = \pi$ in Eq.~(\ref{theta_2}).
Under these numerical parameters, we obtain vortex structures shown in Figs.~\ref{op_3_2}(a) and \ref{op_3_2}(b).
We show vortex structures at top and bottom surface and the center cross section in the superconductor.
In Fig.~\ref{op_3_2}(a), we see that two vortices around $x \sim 20\xi_0$ tilt toward the $x$-axis around $x \sim 10\xi_0$ and $30\xi_0$, while, two vortices around $x \sim 10\xi_0$ and $30\xi_0$, where the magnetic field $(\mbox{\boldmath $H$}_{\rm ext})_y/(\Phi_0/\xi_0^2)= \pm 0.15$, are parallel or antiparallel to $y$-axis.
So, these vortices have quantum fluxes parallel to the direction of $y$-axis. 
Compare to the previous model, sheilding the field along the $y$-axis costs equal energy with shielding the field along the $z$-axis.
Then, vortices parallel to the direction of $y$-axis also appear in this system.

\begin{figure}[t]
 \centering
 \includegraphics[scale=0.27]{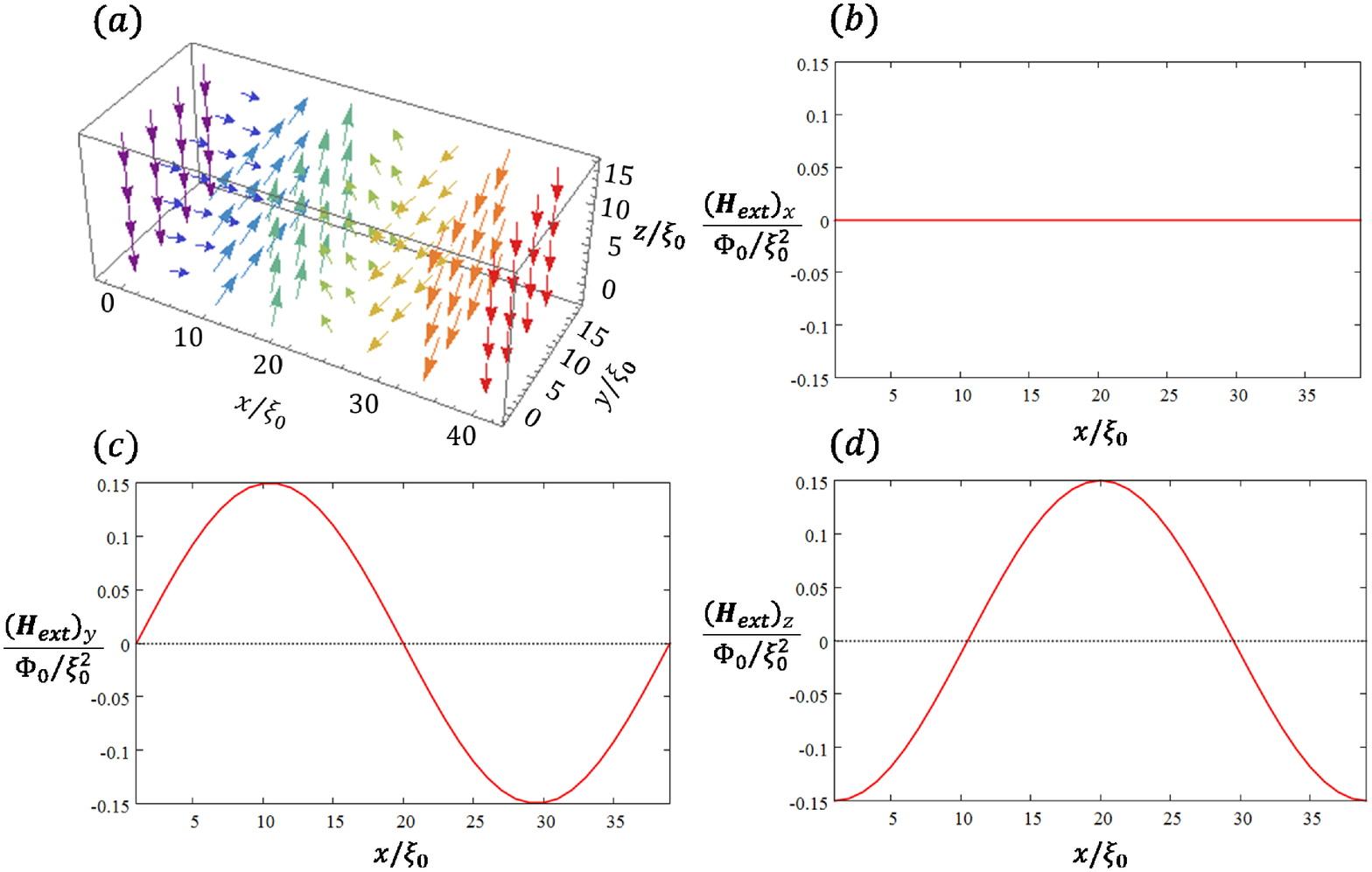}
 \caption{(a) Distributions of the magnetic field from the chiral helimagnet, (b) $x$-component of the magnetic field, (c) $y$-component of the magnetic field, (d) $z$-component of the magnetic field. The amplitude of the helical magnetic field $H_0/(\Phi_0/\xi_0^2) = 0.15$ and the applied magnetic field $H_{\rm appl}/(\Phi_0/\xi_0^2) = 0.00$ for $\phi = \pi$ in Eq.~(\ref{theta_2}). The ratio between the Dzyaloshinsky-Moriya interaction and the ferromagnetic interaction is $|\mbox{\boldmath $D$}|/J=-0.16$.}
 \label{Fig_field_3}
\end{figure}

\begin{figure*}[htbp]
 \centering
 \includegraphics[scale=0.3]{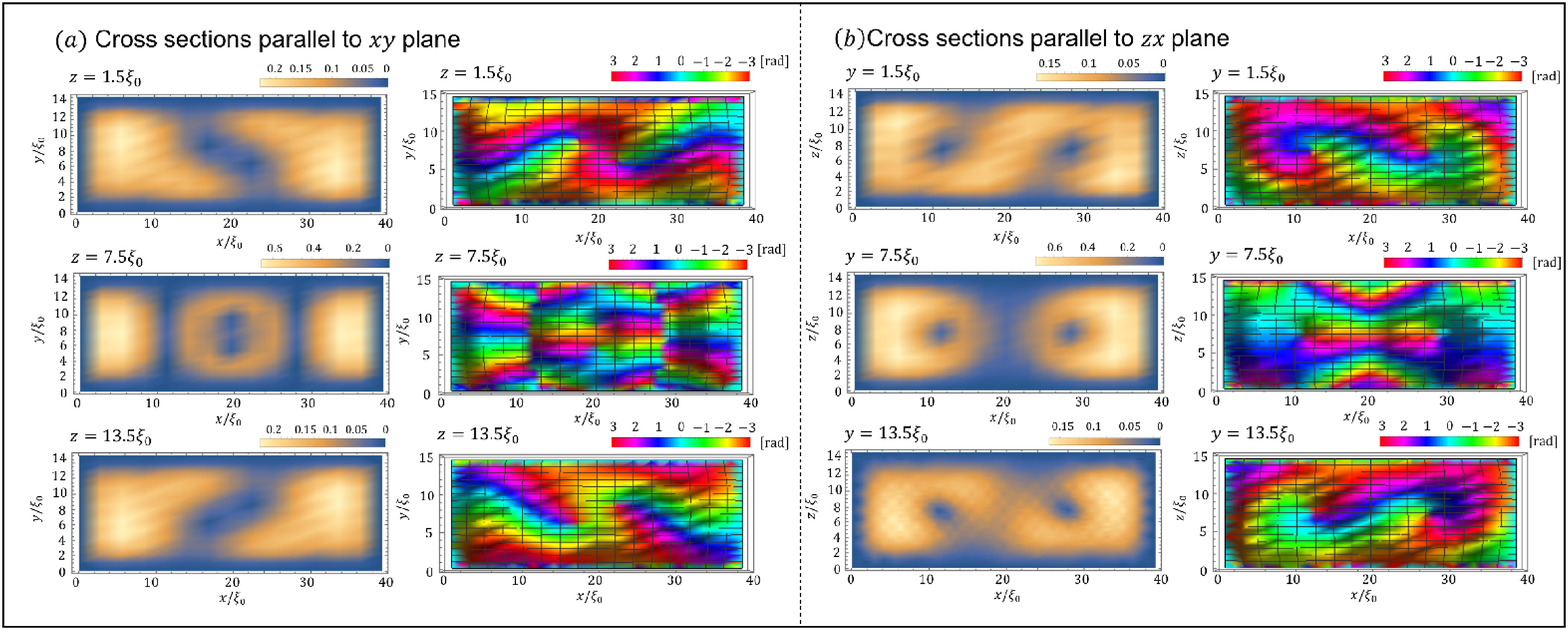}
 \caption{Distributions of the order parameter and phases of the order parameter in the cross sections parallel to (a) $xy$-planes and (b) $zx$-planes. The amplitude of the helical magnetic field $H_0/(\Phi_0/\xi_0^2) = 0.15$ and the applied magnetic field $H_{\rm appl}/(\Phi_0/\xi_0^2) = 0.00$. }
 \label{op_4_2}
\end{figure*}

Next, we examine how the chirality of the helical rotation of magnetic field affects the vortex structures, i.e. we examine difference between vortex structures under right- and left-handed helical magnetic field.
In order to change the direction of the rotation, we take the opposite DM vector, $|\mbox{\boldmath $D$}|/J=0.16$ and $\mbox{\boldmath $D$}$ is antiparallel to the $x$-axis.
The helical magnetic field for this DM vector is shown in Fig.~\ref{Fig_field_3}.
The rotation of the helical magnetic field in Fig.~\ref{Fig_field_3} is opposite to that in Fig.~\ref{Fig_field_2}.
We solve the Ginzburg-Landau equations using this helical magnetic field.
We take the numerical parameter, $\kappa = 5,~T=0.3T_c$, and the magnetic field, $H_0/(\Phi_0/\xi_0^2) = 0.15$ and $H_{\rm appl}/(\Phi_0/\xi_0^2) = 0.0$.
The system size is $1.0L'\xi_0 \times 15\xi_0 \times 15\xi_0$, which is the same system size as Fig.~\ref{op_3_2}.
Under these conditions, we obtain vortex structures shown in Figs.~\ref{op_4_2}(a) and \ref{op_4_2}(b).
Comparing between Figs.~\ref{op_3_2}(a) and \ref{op_4_2}(a), directions of vortices are completely opposite.
Then, the direction of tilt of vortices depends on the chirality of the helical magnetic field.

Experimentally, these vortex structures may appear in superconductor / chiral helimagnet hybrid structures.
For example, our model may be equivalent to the system in which a small superconductor is surrounded by a large chiral helimagnet.
On the other hand, in the chiral helimagnet / superconductor bilayer system, when the superconductor is thin, only the perpendicular component of the magnetic structure is effective.
Then, our previous work on the two-dimensional superconductor is applicable to such bilayer systems\cite{Fukui_SUST,Fukui_JPSJ}.

Finally, we discuss the movement of vortices under the external current.
When the external current is applied to vortex structures in Figs.~\ref{op_1_2}, \ref{op_2_2}, \ref{op_3_2}, and \ref{op_4_2}, along the helical axis, vortices easily move perpendicular to the $x$-axis.
On the other hand, when the external current flows along the direction of the $y$-axis, vortices move in the surface parallel to the $zx$-plane.
If the $z$-component of vortex direction is not zero, the vortex moves toward the $x$-axis.
But the distribution of the helical magnetic field varies along the helical axis ($x$-axis) spatially.
So, the interaction between the vortex and the magnetic field changes.
Then, motion of vortices are obstructed by this interaction, which leads to the increase of the critical current. 

\section{Summary}
We have investigated vortex structures in three-dimensional superconductor under the helical magnetic field from the chiral helimagnet numerically.
We have obtained distributions of the order parameter using the three-dimensional Ginzburg-Landau equations.
When two vortices appear in one magnetic field region $(\mbox{\boldmath $H$}_{\rm ext})_z > 0$, vortices tilt toward the $x$-axis, or the helical axis in spite of $(\mbox{\boldmath $H$}_{\rm ext})_x = 0$.
This configuration may come from the interaction between vortices and between the vortex and the boundary in the system.
It is confirmed that when the rotation of the helical magnetic field is reversed, directions of tilt of vortices are also reversed.
These vortex structures do not occur under the uniform magnetic field.
Detailed discussion of stability of these vortex states need more simulations.

We only consider superconductors under the helical magnetic field, but without the uniform magnetic field and the external current.
Under the uniform magnetic field, the magnetic structure in the chiral helimagnet changes into the chiral soliton lattice.
In the microscopic system, uniform magnetic field decreases the number of solitons in the chiral soliton lattice discretely\cite{CHM_Cr_3}.
Then, we expect that this discrete change of the soliton affects the structures of vortices.
If the external current flows in this system, two vortices that tilt toward the helical axis move uniquely because of the complicated distributions of currents in the superconductor.
Investigations and discussions about these phenomena are future works.


\appendix
 \section{Coefficients for the finite element method}
In this appendix, we give coefficients in Eqs. (\ref{finite_element_3d}), (\ref{e1})-(\ref{e5}) for the finite element method.
First, $a_i, b_i, c_i,$ and $d_i$ in Eqs. (\ref{finite_element_3d}) are defined as,
\begin{eqnarray}
 a_i &=& \epsilon_i \left\{ x_j(y_kz_l - y_lz_k) + x_k(y_lz_j - y_jz_l) \right. \nonumber \\
      & & \left. + x_l(y_jz_k - y_kz_j) \right\}, \label{a_i} \\
 b_i &=& \epsilon_i \left\{  y_j(z_l-z_k) + y_k(z_j-z_l) \right. \nonumber \\
       & & \left. + y_l(z_k-z_j) \right\}, \label{b_i}  \\
 c_i &=& \epsilon_i \left\{  z_j(x_l-x_k) + z_k(x_j-x_l) \right. \nonumber \\
       & & \left. + z_l(x_k-x_j) \right\} ,\label{c_i}  \\
 d_i &=& \epsilon_i \left\{  x_j(y_l-y_k) + x_k(y_j-y_l) \right. \nonumber \\
       & & \left. + x_l(y_k-y_j) \right\}, \label{d_i}       
\end{eqnarray}
where $x_i,~y_i,$ and $z_i~~(i=1,2,3,4)$ are coordinates for nodes in Fig. \ref{finite_element_3d}(b) and $\epsilon_i = 1~~(i=1,3)$ or $\epsilon_i = -1~~(i=2,4)$.
$(i,j,k,l)$ is a cyclic permutation of (1,2,3,4).

Then, we show coefficients in Eqs. (\ref{e1})-(\ref{e5}).
They are given as,
\begin{eqnarray}
 P_{ij}(\{\mbox{\boldmath $A$}\}) &=& \sum_{\alpha} K_{ij}^{\alpha\alpha} + \sum_{i_1,i_2} I_{i_1i_2ij} \sum_{\alpha} A_{i_1\alpha}^e A_{i_2\alpha}^e - \frac{1}{\xi(T)^2} I_{ij}, \nonumber \\ \label{p} \\
 P_{ij}^{2R}(\{\psi\}) &=& \frac{1}{\xi(T)^2} \sum_{i_1i_2} I_{i_1i_2ij} (3{\rm Re}\psi_{i_1}^e {\rm Re}\psi_{i_2}^e + {\rm Im}\psi_{i_1}^e {\rm Im}\psi_{i_2}^e), \nonumber \\ \label{p2r} \\
 P_{ij}^{2I}(\{\psi\})  &=& \frac{1}{\xi(T)^2} \sum_{i_1i_2} I_{i_1i_2ij} ({\rm Re}\psi_{i_1}^e {\rm Re}\psi_{i_2}^e + 3{\rm Im}\psi_{i_1}^e {\rm Im}\psi_{i_2}^e), \nonumber \\ \label{p2i} \\
 Q_{ij}(\{\mbox{\boldmath $A$}\}) &=& \sum_{i_1} \sum_{\alpha} (J_{ji_1i}^{\alpha} - J_{ii_1j}^{\alpha}) A_{i_1}^{\alpha}, \label{q} \\
 Q_{ij}^2(\{\psi\})     &=& \frac{2}{\xi(T)^2} \sum_{i_1i_2} I_{i_1i_2ij} {\rm Re} \psi_{i_1}^e {\rm Im} \psi_{i_2}^e, \label{q2} \\
 R_{ij}(\{\psi\})         &=& \kappa^2 \xi(T)^2 \sum_{\alpha} K_{ij}^{\alpha\alpha} + \sum_{i_1i_2} I_{i_1i_2ij} \psi_{i_1}^{e\ast} \psi_{i_2}^e, \label{r} \\
 T_{i}^{\alpha}(\{\psi\})         &=& \sum_{i_1i_2} J_{i_1i_2i}^{\alpha} {\rm Im} (\psi_{i_1}^{e\ast} \psi_{i_2}^e), \label{t} \\
 V_{i}^{R}(\{\psi\})               &=& \frac{2}{\xi(T)^2} \sum_{i_1i_2i_3} I_{i_1i_2i_3i} {\rm Re}(\psi_{i_1}^e \psi_{i_2}^{e\ast}) {\rm Re} \psi_{i_3}^e, \label{vr} \\
 V_{i}^{I}(\{\psi\})                &=& \frac{2}{\xi(T)^2} \sum_{i_1i_2i_3} I_{i_1i_2i_3i} {\rm Re}(\psi_{i_1}^e \psi_{i_2}^{e\ast}) {\rm Im} \psi_{i_3}^e, \label{vi} \\
 S_{ij}^{\alpha\beta} &=& \kappa^2 \xi(T)^2 (K_{ij}^{\alpha\beta} - K_{ij}^{\beta\alpha}), \label{s} \\
 U_{i}^{\alpha}        &=& \kappa^2 \xi(T)^2 \frac{2\pi}{\Phi_0} \left( J_i^{\beta}H_{\gamma} - J_i^{\gamma}H_{\beta}  \right). \label{u} 
\end{eqnarray} 
Here, $\alpha$, and  $\beta=x,y,z$.
In Eq.~(\ref{u}), $(\alpha, \beta, \gamma)$ is a cyclic permutation of $(x,y,z)$.
Their coefficients are represented by integrals $I_{ij}$, $I_{i_1i_2i_3}$, $I_{i_1i_2i_3i_4}$, $J_i^{x_i}$, $J_{i_1i_2i_3}^{x_i}$, and $K_{ij}^{x_ix_j}$.
These integrals are given by,
\begin{eqnarray}
 I_{ij} &=& \int_{V_e} N_i^e N_j^e dV, \label{int_i2} \\
 I_{i_1i_2i_3} &=& \int_{V_e} N_{i_1}^e N_{i_2}^e N_{i_3}^e dV, \label{int_i3} \\
 I_{i_1i_2i_3i_4} &=& \int_{V_e} N_{i_1}^e N_{i_2}^e N_{i_3}^e N_{i_4}^e dV, \label{int_i4} \\
 J_{i}^{x_i} &=& \int_{V_e} \frac{\partial N_i^e}{\partial x_i} dV, \label{int_j1} \\
 J_{i_1i_2i_3}^{x_i} &=& \int_{V_e} \frac{\partial N_{i_1}^e}{\partial x_i}N_{i_2}^e N_{i_3}^e dV, \label{int_j3} \\
 K_{i_1i_2}^{x_ix_j} &=& \int_{V_e} \frac{\partial N_{i_1}^e}{\partial x_i} \frac{\partial N_{i_2}^e}{\partial x_j} dV. \label{int_k}
\end{eqnarray}


\begin{thebibliography}{99}
 \bibitem{DM_1}
  I. Dzyaloshinsky, J. Phys. Chem. Solids {\bf 4}, 241 (1958).
 \bibitem{DM_2}
  T. Moriya, Phys. Rev. {\bf 120}, 91 (1960).
 \bibitem{CHM_Cr_1}
  N. J. Ghimire, M. A. McGuire, D. S. Parker, B. Sipos, S. Tang, J.-Q. Yan, B. C. Sales, and D. Mandrus, Phys. Rev. B {\bf 87}, 104403 (2013).
 \bibitem{CHM_Cr_2}
  K. Tsuruta, M. Mito, H. Deguchi, J. Kishine, Y. Kousaka, J. Akimitsu, and K. Inoue, Phys. Rev. B {\bf 93}, 104402 (2016).
 \bibitem{CHM_Cr_3}
  Y. Togawa, Y. Kousaka, K. Inoue, and J. Kishine, J. Phys. Soc. Jpn. {\bf 85}, 11201 (2016).
 \bibitem{CHM_Cs_1}
  H. Ohsumi, A. Tokuda, S. Takeshita, M. Takata, M. Suzuki, N. Kawamura, Y. Kousaka, J. Akimitsu, and T. Arima, Angew. Chem., Int. Ed. {\bf 52}, 8718 (2013).
 \bibitem{CHM_Cs_2}
  Y. Kousaka, T. Koyama, K. Ohishim, K. Kakurai, V. Hutanu, H. Ohsumi, T. Arima, A. Tokura, M. Suzuki, N. Kawamura, A. Nakao, T. Hanashima, J. Suzuki, J. Campo, Y. Miyamoto, A. Sera, K. Inoue, and J. Akimitsu, Phys. Rev. Material {\bf 1}, 071402(R) (2017).
 \bibitem{CHM_Yb_1}
  S. Ohara, S. Fukuta, K. Ohta, H. Kono, T. Yamashita, Y. Matsumoto, and J. Yamaura, JPS Conf. Proc. {\bf 3}, 017016 (2014).
 \bibitem{CHM_Yb_2}
  T. Matsuura, Y. Kita, K. Kubo, Y. Yoshikawa, S. Michimura, T. Inami, Y. Kousaka, K. Inoue, and S. Ohara, J. Phys. Soc. Jpn. {\bf 86}, 124702 (2017).
 \bibitem{CHM_theory_1}
  M. Shinozaki, S. Hoshino, Y. Masaki, J. Kishine, and Y. Kato, J. Phys. Soc. Jpn {\bf 85}, 074710 (2016).
 \bibitem{Kishine_theory}
  J. Kishine and A. S. Ovchinnnikov, Solid State Phys. {\bf 66}, 1 (2015).
 \bibitem{Kishine_1}
  J. Kishine, K. Inoue, and Y. Yoshida, Prog. Theor. Phys. Suppl. {\bf 159}, 82 (2005).
 \bibitem{Togawa_PRL_1}
  Y. Togawa, T. Koyama, K. Takayanagi, S. Mori, Y. Kousaka, J. Akimitsu, S. Nishihara, K. Inoue, A. S. Ovchinnikov, and J. Kishine, Phys. Rev. Lett. {\bf 108}, 107202 (2012).
 \bibitem{Togawa_PRL_2}
  Y. Togawa, Y. Kousaka, S. Nishihara, K. Inoue, J. Akimitsu, A. S. Ovchinnikov, and J. Kishine, Phys. Rev. Lett. {\bf 111}, 197204 (2013)
 \bibitem{MCh_Dh}
  G. L. J. A. Rikken and E. Raupach, Nature {\bf 390}, 493 (1997)
 \bibitem{spin_current}
  K. Tokushuku, J. Kishine, and M. Ogata, J. Phys. Soc. Jpn. {\bf 86}, 124701 (2017).
 \bibitem{BKT}
  I. Proskurin, A. S. Ovchinnikov, and J. Kishine, J. Phys.: Conf. Ser. {\bf 903}, 012062 (2017).
 \bibitem{Abrikosov} 
  A. Abrikosov, Sov. Phys. JETP {\bf 5}, 1774 (1957).
 \bibitem{Hess}
  H. F. Hess, R. B. Robinson, R. C. Dynes, J. M. Valles, Jr., and J. V. Waszczak, Phys. Rev. Lett. {\bf 62}, 214 (1989).
 \bibitem{FM_SC_Hybrid}
  I. F. Lyuksyutov and V. L. Pokrovsky, Adv. Phys. {\bf 54}, 1 (2005).
 \bibitem{FM_SC_FSB_ex}
  M. Lange, M. J. Van Bael, V. V. Moshchalkov, and Y. Bruynseraede, Appl. Phys. Lett., {\bf 81}, 322 (2002).
 \bibitem{FM_SC_FSB_theory}
  M. A. Kayali and V. L. Pokrovsky, Phys. Rev. B {\bf 69}, 132501 (2004).  
 \bibitem{Fukui_SUST}
  S. Fukui, M. Kato and Y. Togawa, Supercond. Sci. Technol. {\bf 29}, 125008 (2016)
 \bibitem{Fukui_JPSJ}
  S. Fukui, M. Kato, Y. Togawa, and O. Sato, submitted to J. Phys. Soc. Jpn. 
 \bibitem{Fukui_3d_proc}
  S. Fukui, M. Kato, Y. Togawa, and O. Sato, submitted to J. Phys.: Conf. Ser.
 \bibitem{DM_Cr}
  B. J. Chapman, A. C. Bornstein, N. J. Ghimire, D. Mandrus, and M. Lee, Appl. Phys. Lett. {\bf 105}, 072405 (2014).
 \end{thebibliography}
\end{document}